\documentclass[twocolumn,english,aps,prl,superscriptaddress]{revtex4}
\pdfoutput=1
\usepackage{graphicx}
\usepackage{amssymb,amsmath}
\usepackage{textcomp} 
\usepackage[bold]{hhtensor} 

\newcommand{\tocless}[2]{\bgroup\let\addcontentsline=\nocontentsline#1{#2}\egroup}

\begin{document}
\title{Synchronization and Phase Noise Reduction in Micromechanical Oscillators Arrays Coupled through Light}

\author{Mian Zhang}
\affiliation{School of Electrical and Computer Engineering, Cornell University, Ithaca, New York 14853, USA.}

\author{Shreyas Shah}
\affiliation{School of Electrical and Computer Engineering, Cornell University, Ithaca, New York 14853, USA.}

\author{Jaime Cardenas}
\affiliation{School of Electrical and Computer Engineering, Cornell University, Ithaca, New York 14853, USA.}

\author{Michal Lipson}
\affiliation{School of Electrical and Computer Engineering, Cornell University, Ithaca, New York 14853, USA.}
\affiliation{Kavli Institute at Cornell for Nanoscale Science, Ithaca, New York 14853, USA.}
\date{\today}



\begin{abstract}
Synchronization of many coupled oscillators is widely found in nature and has the potential to revolutionize timing technologies. Here we demonstrate synchronization in arrays of silicon nitride micromechanical oscillators coupled in an all-to-all configuration purely through an optical radiation field. We show that the phase noise of the synchronized oscillators can be improved by almost 10 dB below the phase noise limit for each individual oscillator. These results open a practical route towards synchronized oscillator networks.
\end{abstract}
\maketitle

\date{\today}
\newcommand{\nocontentsline}[3]{}

Nano- and micromechanical oscillator arrays have the potential to enable high power and low noise integrated frequency sources that play a key role in sensing and the essential time keeping of modern technology \cite{craighead_nanoelectromechanical_2000,zhang_rapid_2006,shim_synchronized_2007,bargatin_large-scale_2012,nguyen_mems_2007}. The challenge with building scalable oscillator arrays is that micromechanical oscillators fabricated on a chip fundamentally have a spread of mechanical frequencies due to unavoidable statistical variations in the fabrication process \cite{bargatin_large-scale_2012,matheny_phase_2014,bagheri_photonic_2013,zhang_synchronization_2012,agrawal_observation_2013}. This dispersion in mechanical frequencies has a detrimental effect on the coherent operation in arrays of micromechanical oscillators. Here we show that arrays consisting of three, four and seven dissimilar microscale optomechanical oscillators can be synchronized to oscillate in unison coupled purely through a common optical cavity field using less than a milliwatt of optical power. We further demonstrate that the phase noise of the oscillation signal can be reduced by a factor of $N$ below the thermomechanical phase noise limit of each individual oscillator as $N$ oscillators are synchronized, in agreement with theoretical predictions \cite{pikovsky_synchronization:_2003,chang_phase_1997}. The highly efficient, low loss and controllable nature of light mediated coupling could put large scale nano- and micromechanical oscillator networks in practice \cite{heinrich_collective_2011,holmes_synchronization_2012,shah_master-slave_2015,safavi-naeini_two-dimensional_2014,aspelmeyer_cavity_2014,botter_optical_2013, shkarin_optically_2014}.

Synchronization is a ubiquitous phenomenon found in coupled oscillator systems \cite{pikovsky_synchronization:_2003,strogatz_sync:_2003}. Heart beat is a result of synchronized motion of pace maker cells \cite{peskin_mathematical_1975}, circadian rhythm arises because of coordinated body physiology \cite{reppert_coordination_2002} and global positioning system relies on synchronized operation of clocks. On the nanoscale, synchronization has been experimentally demonstrated in nanomechanical systems coupled through mechanical connections \cite{shim_synchronized_2007}, electrical capacitors \cite{agrawal_observation_2013}, off-chip connections \cite{matheny_phase_2014} and through an optical cavity \cite{bagheri_photonic_2013,zhang_synchronization_2012}. However, these demonstrations were limited to only two oscillators. Achieving synchronization in large micromechanical oscillator networks requires scalable oscillator units and efficient and controllable coupling mechanisms \cite{heinrich_collective_2011,holmes_synchronization_2012,lauter_pattern_2015}.
\begin{figure}
   \centering
   \includegraphics[angle=0,width=0.45\textwidth]{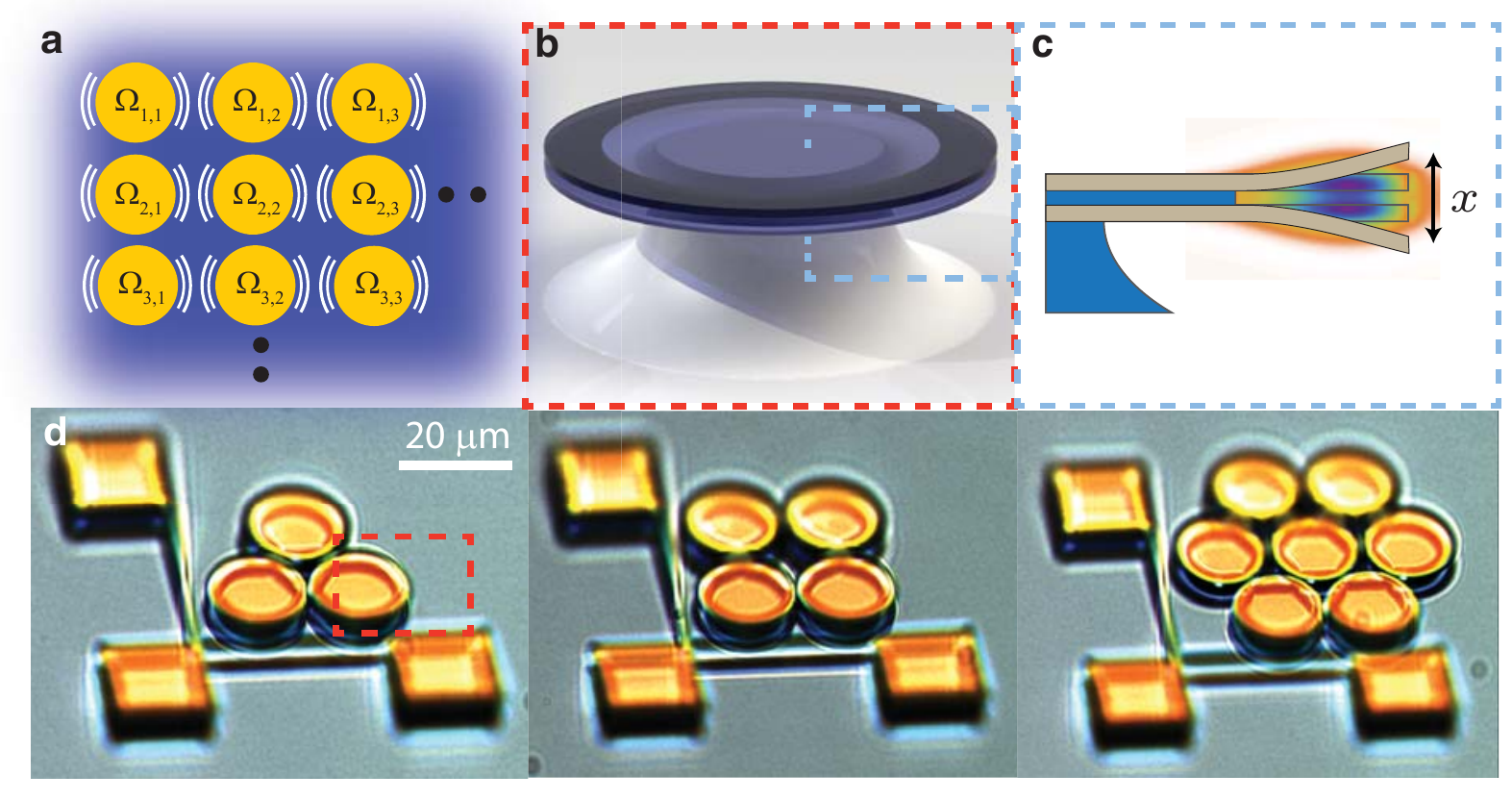}

 \caption{\textbf{Concept and devices} (a) Concept of mediating coupling between mechanical oscillators (yellow) through a global optical field (blue). The optical field provides energy for each mechanical oscillator to vibrate at their natural frequencies $\Omega_{i,j}$ and also provide coupling between each mechanical oscillator forming an all-to-all coupling topology. When the optical coupling is strong, the oscillators synchronize and vibrate at a common frequency. (b) A schematic of each individual double-disk. The edges are partly suspended to allow for mechanical vibration  (c) Cross section of a double-disk showing the mechanical and the optical mode shapes. (d) Optical microscope images of coupled optomechanical double-disk oscillator arrays. The oscillators are mechanically separated by a narrow gap ($\sim 150$ nm) and coupled solely through the optical evanescent field. The squares and strings are support structures for tapered optical fibers.}
  \label{fig:figure1}
 \end{figure}

Here we experimentally demonstrate that arrays of free running micromechanical oscillators can be synchronized when coupled purely through a common electromagnetic field as predicted by theories \cite{heinrich_collective_2011,holmes_synchronization_2012}. A conceptual view of an array of mechanical resonators coupled through light is illustrated in Figure \ref{fig:figure1}a. Each optomechanical oscillator (OMO) possesses a slightly different frequency of mechanical oscillation ($\Omega_i$) and is only connected through a common optical field (blue background). When a continuous wave laser is coupled to a common electromagnetic field mode spanning several micromechanical oscillators, the light can provide both the drive for self-sustaining oscillations and the necessary coupling between the individual oscillators for synchronization through optical forces. When the laser power is just above the self-sustaining oscillation threshold of the mechanical oscillators, they are expected to vibrate at their natural frequencies $\Omega_i$. When the laser power is high so that the optically mediated coupling is strong enough to overcome the difference in $\Omega_i$, the mechanical oscillators can reach synchronization.

The effective coupling between the mechanical resonators can be visualized through the following equation 
\begin{equation}
\begin{aligned}
\ddot{x_i}+\Gamma_i\dot{x_i} +\Omega_i^2 x_i=F^{(i)}_{\textnormal{opt}},
\\
F^{(i)}_{\textnormal{opt}}\propto |b(x_i,...,x_j)|^2
\end{aligned}
\end{equation}
where $x_i$,$\Gamma_i$,$\Omega_i$ are the mechanical displacement, damping and the mechanical frequency of the $i$th OMO and $b(x_i,...,x_j)$ is the amplitude of the coupled optical supermode that spatially spans all cavities in the array. It is clear from the equation above that the optical force ($F_{\rm{opt}}$) depends on the energy stored in the optical supermode which is affected by the displacement of each individual cavity. Therefore the optical field provides an effective nonlinear mechanical coupling between the different oscillators that form the basis for synchronization \cite{zhang_synchronization_2012,bagheri_photonic_2013,holmes_synchronization_2012}. The onset of synchronization, which intrinsically relies on nonlinearity \cite{cross_synchronization_2004}, could therefore be captured as ($F_{\rm{opt}}$) is increased through increasing the optical driving power \cite{bagheri_photonic_2013}.

\begin{figure}
   \centering
   \includegraphics[angle=90,width=0.5\textwidth]{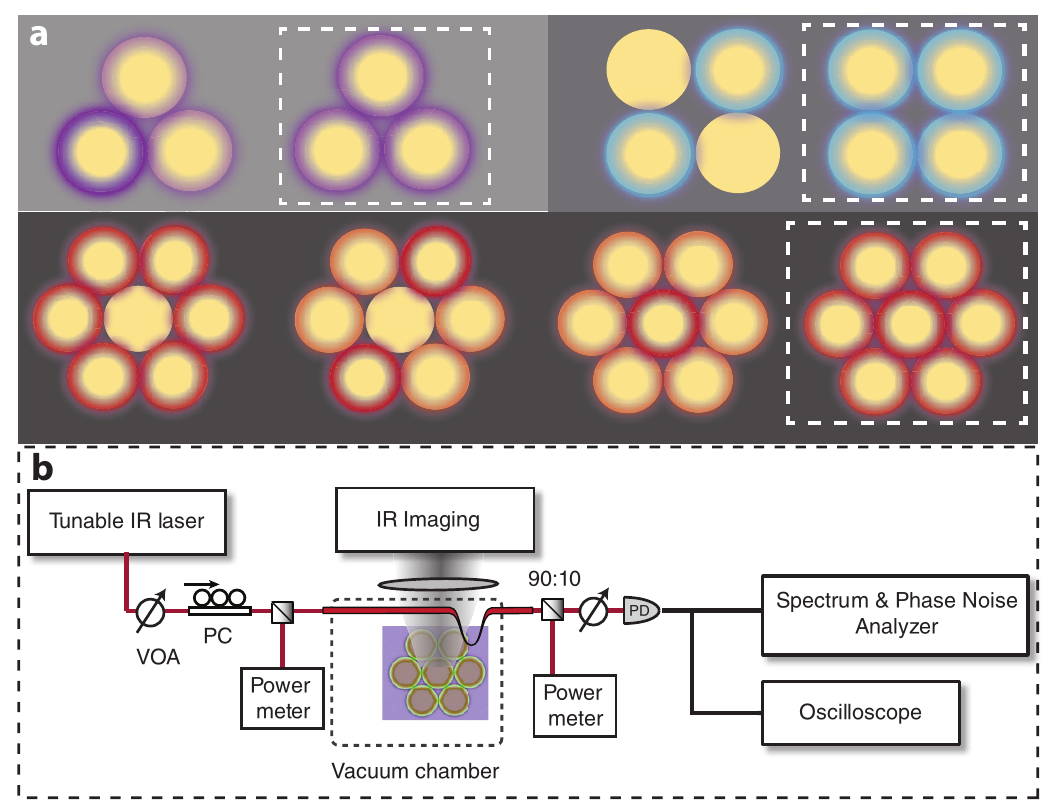}

 \caption{\textbf{Experimental configuration} (a) Optical supermodes spatial structures. The colored halos show where the optical cavity field resides for different types of arrays. The more opaque colors illustrate higher cavity field intensities when compared to the rest of the cavities. The supermodes that spatially span over all cavities with equal intensities are identified by dashed lines.  (b) Experimental setup. The coupled optomechanical oscillator array is placed in a vacuum chamber and excited by a tunable infra-red (IR) camera through a tapered optical fiber. The optical power and polarization are controlled by an variable optical attenuator (VOA) and a fiber polarization controller (PC). The optical transmission is detected by an amplified photodiode (PD) and analyzed by an oscilloscope and a spectrum analyzer.}
  \label{fig:figure2}
 \end{figure}
 
The individual oscillator we use is a double-disk OMO (Figures \ref{fig:figure1}b and \ref{fig:figure1}c) composed of two free-standing silicon nitride circular edges that support high quality ($Q$) factor optical and mechanical modes \cite{zhang_eliminating_2014,lin_mechanical_2009}. The co-localized modes shown in Figure \ref{fig:figure1}c lead to a strong coupling between the optical and the mechanical degree of freedom. When the cavity is excited by a continuous wave laser above the oscillation threshold, the free-standing edges oscillate coherently and modulate the laser producing a radio frequency (RF) tone at the mechanical frequency of the vibrating edges. Fabrication variation causes the mechanical frequency of different OMOs in our arrays to spread around $\pm 1$ MHz centered at 132.5 MHz \cite{seeSI}. 

\begin{figure*}
   \centering
   \includegraphics[angle=0,width=1\textwidth]{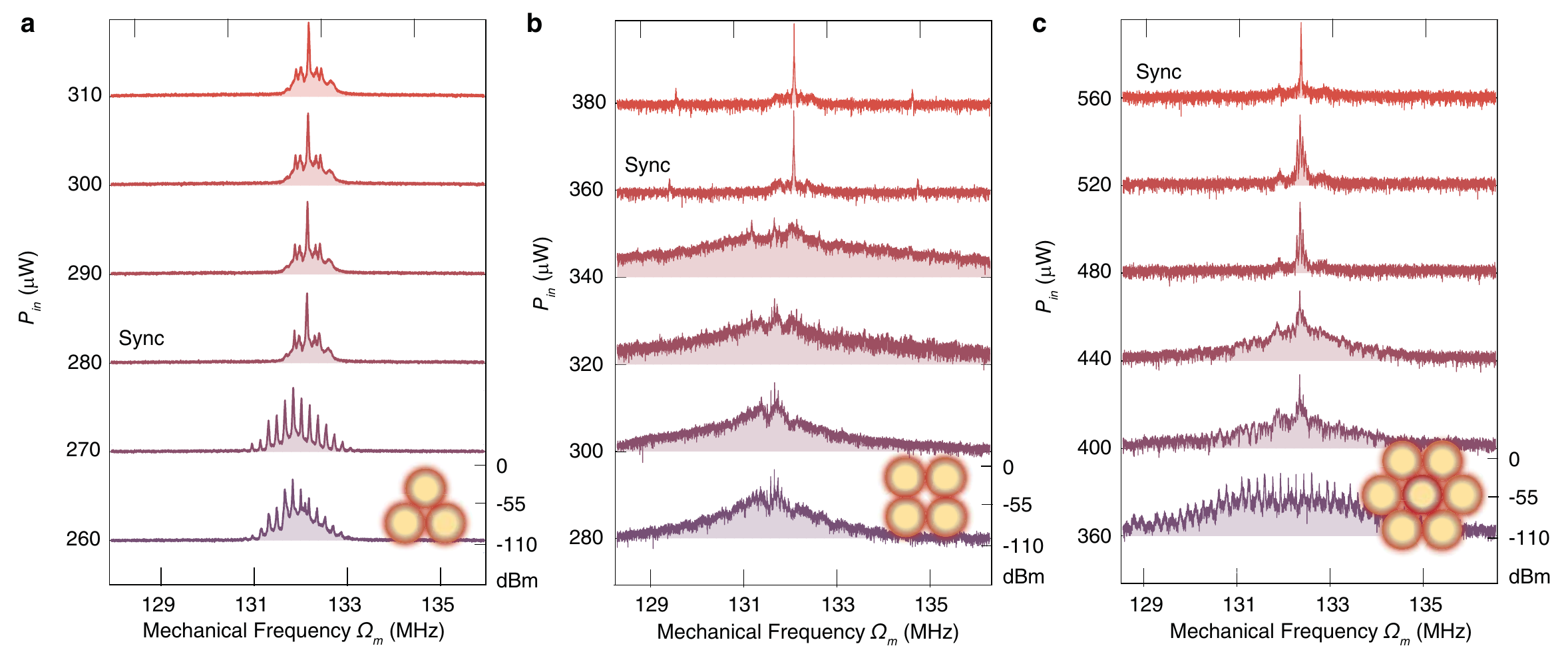}

 \caption{\textbf{Synchronization in arrays of OMOs} Optical power spectrum of three- (a) four- (b) and seven- (c) OMO system as the input optical power increases. The vertical scale is from -110 dBm to 0 dBm for each trace. Synchronization is characterized by the sudden noise floor drop and the emergence of a single frequency in the optical power spectrum as indicated in the graphs. The disorder in natural mechanical frequencies and incoherent dynamics before the onset of synchronization is evident from the RF peaks and the broad noise floor. The seven resonator array (c) shows multiple changes of noise shapes before eventually synchronize indicating the presence of multiple oscillation states as a result of many OMOs.}
  \label{fig:figure3}
 \end{figure*}
 
We fabricate micromechanical oscillator arrays with double-disk OMOs that are optically coupled through the evanescent field. The OMOs are physically separated by a narrow gap ($\sim 150$ nm) which precludes any mechanical connections while the optical evanescent field can still propagate through the gap. Mechanical coupling through the substrate connection is negligible as the mechanical mode we excite is a high $Q$ mode that is well isolated from the substrate \cite{zhang_eliminating_2014}.

We excite the optical supermode that spatially span over all cavities to ensure that there is optical coupling among all cavities (Fig. \ref{fig:figure2}a dashed boxes). The strong optical coupling between the optical modes of each individual cavity $a_i$ leads to the formation of optical supermodes $b_m$ that have different optical frequencies and spatial geometries \cite{seeSI}. Figure \ref{fig:figure2}a illustrates the spatial intensity profile of different optical supermodes $b_m$ when the optical resonant frequency of individual cavity $\omega_i$ is identical. The higher intensity regions are illustrated by higher opacity of the halos around the cavities. We position a tapered optical fiber to the close proximity of one OMO in the arrays to couple light to the spatial evenly distributed optical modes (dashed lines in Fig. \ref{fig:figure2}a) while using an infrared (IR)-camera to monitor the scattered intensity from the arrays making sure all OMOs are excited. We monitor the transmission through the tapered fiber by an amplified photodiode and feed the electrical signal to a spectrum analyzer.

We show the onset of synchronization by increasing the excitation laser power which effectively increases the coupling between the OMOs. The laser wavelength is blue detuned relative to the resonance of the optical supermode that evenly spans all the OMOs (Fig. \ref{fig:figure2}a dashed boxes) enabling optomechanical amplification. In the three coupled OMO array, as the laser power increases well beyond the oscillation threshold for each individual oscillator, the radio frequency (RF) spectrum of the OMOs show many strong oscillation peaks and a broad noise floor (Fig. \ref{fig:figure3}a). The distinct oscillation peaks form because $\Omega_i$ is different for each OMO and they beat to generate many RF tones \cite{hossein-zadeh_observation_2008,huang_internal_2012}. The increase in the noise floor is likely due to finite interaction between the mechanical modes mediated by the optical field but not yet strong enough to transition into a locked state \cite{bagheri_photonic_2013,okawachi_octave-spanning_2011,delhaye_self-injection_2014}. As the laser power further increases to $P_{\rm{in}} = 280~\mu$W, the onset of synchronization (Fig. \ref{fig:figure3}a) is evident as the peaks on the RF spectrum merge into a single large peak and the noise floor is reduced. The much weaker sidebands around the main oscillation signal are due to the much weaker oscillatory motion induced by thermal force displacing the OMOs from the synchronized state \cite{bagheri_photonic_2013}. In four and seven coupled OMO arrays, similar to the three-cavity system, we observe beating between different mechanical modes and a broad noise floor when the optical power is below the synchronization threshold. As the laser power is increased, a single oscillation peak appears accompanied with a sudden drop in the noise floor signifying the onset of synchronization (Fig. \ref{fig:figure3}b,c).
\begin{figure*}
     \centering
     \includegraphics[angle=0,width=1\textwidth]{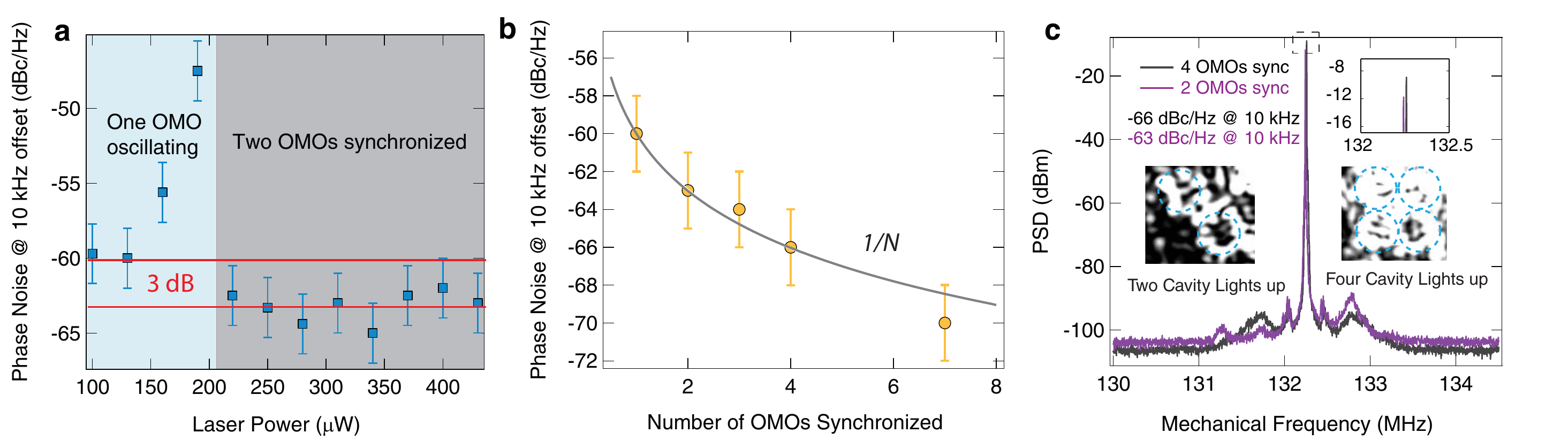}

   \caption{\textbf{Phase noise in synchronized arrays} (a) phase noise in a two-OMO system at 10 kHz carrier offset as a the laser power is increased. The noise increases due to mode competition between possible oscillation states and  then decreases by  $\sim 3$ dB below the noise level of one OMO oscillation state. (b) The phase noise of the synchronized oscillation signal for different sizes of OMO arrays. The grey curve is the phase noise level predicted by theory for near-identical synchronized oscillators (c) Power spectrum of a state where four OMO are oscillating (black) and of a state two OMO oscillating (purple). The phase noise drops by $\sim 3$ dB following the transition. The inset shows a zoom in near the peak and captured IR images for the two and four OMO synchronized state respectively.}
      \label{fig:figure4}
\end{figure*}

We show that in large arrays of OMOs, the phase noise of the synchronized signal can be reduced below the thermomechanical noise limit of an individual OMO by almost 10 dB. The phase noise of the modulated output light is expected to drop as the oscillators are synchronized \cite{matheny_phase_2014,pikovsky_synchronization:_2003,chang_phase_1997}. We measure the phase noise of our oscillators at 10 kHz offset from the carrier oscillation frequency, where the phase noise of our oscillator is dominated by thermomechanical fluctuation \cite{fong_phase_2014,tallur_phase_2010,hossein-zadeh_characterization_2006,seeSI}, a fundamental limit imposed to the mechanical oscillator due to the thermal bath of the environment. In figure \ref{fig:figure4}a, we show the measured phase noise in a $1\times 2$ OMO array \cite{zhang_synchronization_2012}. As shown in figure \ref{fig:figure4}a, the single OMO phase noise \cite{seeSI} at low optical power is $\sim -60$ dBc/Hz and gradually increases as the laser power is increased. The increase of phase noise is due to phase slipping between the two OMOs \cite{matheny_phase_2014}. As the coupling between the OMOs increases with increasing laser power, they synchronize. As expected, we observe the phase noise drops by $\sim 3$ dB as the two OMOs move from one OMO oscillating state to a synchronized oscillation state. Since the oscillators are nearly identical, synchronized oscillations can be viewed as two oscillators operating coherently providing a larger effective mass while not reducing the oscillation frequency \cite{tallur_phase_2010}. In figure \ref{fig:figure4}c, we show the measured phase noise of each large array of oscillators by driving the system at high optical powers at the optimal optical detuning where the phase noise is a minimum \cite{seeSI}. The lowest phase noise measured in each array of different sizes is plotted in figure \ref{fig:figure4}b. The measured phase noise follows the $1/N$ dependence predicted by theory \cite{pikovsky_synchronization:_2003,chang_phase_1997,tallur_phase_2010, cross_improving_2012}.
  
The drop in phase noise can also be used to determine the number of synchronized OMOs in a single array oscillating in different states. We measure the phase noise in the $2\times 2$ array as the oscillators change from a state where only two OMOs are oscillating to a state where all four OMOs are oscillating, as we infer from the light scattering intensities captured on the IR camera. Figure \ref{fig:figure4}c shows the power spectrum of the transmitted light when the laser is tuned from an optical mode that spans two cavities to an optical mode that spans all four cavities (Fig. \ref{fig:figure2}a) while staying at the same optical power. Following the transition, the four OMO oscillation state shows an increase of $\sim 3$ dB in the oscillation signal and $\sim 3$ dB drop in the phase noise. At the same time, all four resonators light up on the IR camera. The drop in phase noise and the change of scattering intensity on the IR camera strongly indicate that the array changes from two to four synchronized oscillators.

In conclusion, we demonstrate synchronization in integrated arrays of micromechanical oscillators coupled through a common optical field. We show the onset of synchronization in the arrays by tracking the emergence of a single oscillation frequency in the optical power spectrum. Synchronization is further corroborated by the drop in phase noise in the oscillator arrays. The reduction of phase noise with oscillator array size and the scalability of our devices could enable low noise and high power integrated frequency sources. Our work paves a path towards large scale monolithically fabricated oscillator networks that have the potential to compete with the performance of bulk resonators and to exhibit rich nonlinear dynamics opening the door to novel metrology, communication and computing techniques \cite{hoppensteadt_synchronization_2001,mari_measures_2013}. 

We acknowledge Richard Rand and Gustavo Wiederhecker for fruitful discussion about our results. The authors gratefully acknowledge support from DARPA for award No. W911NF-11-1-0202. The authors also acknowledge Applied Optronics and from DARPA for award No. W911NF-14-C0113. This work was performed in part at the Cornell NanoScale Facility, a member of the National Nanotechnology Infrastructure Network, which is supported by the National Science Foundation (Grant ECCS-0335765). This work made use of the Cornell Center for Materials Research Facilities supported by the National Science Foundation under Award Number DMR-1120296. The authors also acknowledge Paul McEuen for use of lab facilities.

\vspace{-15pt}
\bibliographystyle{nature}
\bibliography{multisync2}

\onecolumngrid
\setcounter{figure}{0}
\setcounter{table}{0}
\setcounter{equation}{0}
\renewcommand{\theequation}{S\arabic{equation}}
\renewcommand{\thesection}{S\arabic{section}}
\renewcommand{\thesubsection}{\Alph{subsection}}
\renewcommand{\thesubsubsection}{\roman{subsubsection}}
\renewcommand{\thefigure}{S\arabic{figure}}
\renewcommand{\thetable}{S\arabic{table}}



\setcounter{figure}{0}
\setcounter{table}{0}
\setcounter{equation}{0}
\newpage
\section{Supplemental Information}
\subsection{Optical Transmission}

\begin{figure}[!h]
  \centering
  \includegraphics[angle=0,width=0.6\textwidth]{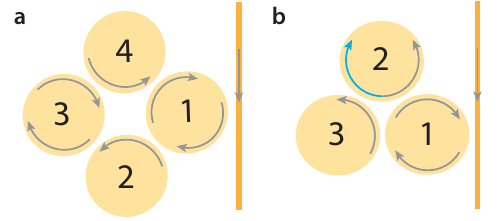}
\caption{Mode excitation geometry (a) In a square lattice, only one direction of the optical mode is excited. (b) In a triangular lattice, the coupling geometry determines that as the light from the waveguide excites the clockwise mode of OMO 1 which in turn excites the counter-clockwise mode of OMO 2 and OMO 3. However, the counter-clockwise mode of OMO 3 also excite the clockwise mode of OMO 2 (blue arrow), which therefore excites all the counter propagating wave of the entire system.}
\label{fig:cwccw}
\end{figure}

The challenge in optically coupling large arrays of mechanical structures is the variation of the optical frequencies of each individual OMO due to variation in the exact fabricated dimensions. This can be in principle compensated by integrating optical tuning mechanisms such as heaters. In our device, we overcome this challenge by ensuring that the optical coupling is large compare to the frequency variation and the individual decay rate of each cavity. The optical coupling strength we designed ($\kappa_{(i,j)}\sim 5$ GHz) is much larger than the optical decay rate of the cavity ($\gamma_i \sim 400$ MHz) and to the cavity optical frequency spread from fabrication variations ($\Delta\omega<2$ GHz). This ensures that the optical modes are strongly coupled which means the light travels between the cavities many times before it is lost via other channels.

We estimate the spatial structure of the optical supermodes using coupled mode theory \cite{Hau84}. In the $2\times 2$ OMO array, the coupled modes can be solved essentially using the same method as described in previous work of two coupled double disk OMOs \cite{zhang_synchronization_2012}. The evolution and coupling of each individual optical mode is described by a matrix where its diagonal entries describe the eigenfrequencies and the off-diagonal entries describe the coupling between these modes. In the case for the three- and seven- OMO arrays arranged in a triangular lattice, we must consider both the clock-wise and counter-clock wise optical modes, as both will be excited through a laser coupled to one direction even without any backscatters (Fig. \ref{fig:cwccw}). 

\begin{figure}
  \centering
  \includegraphics[angle=0,width=1\textwidth]{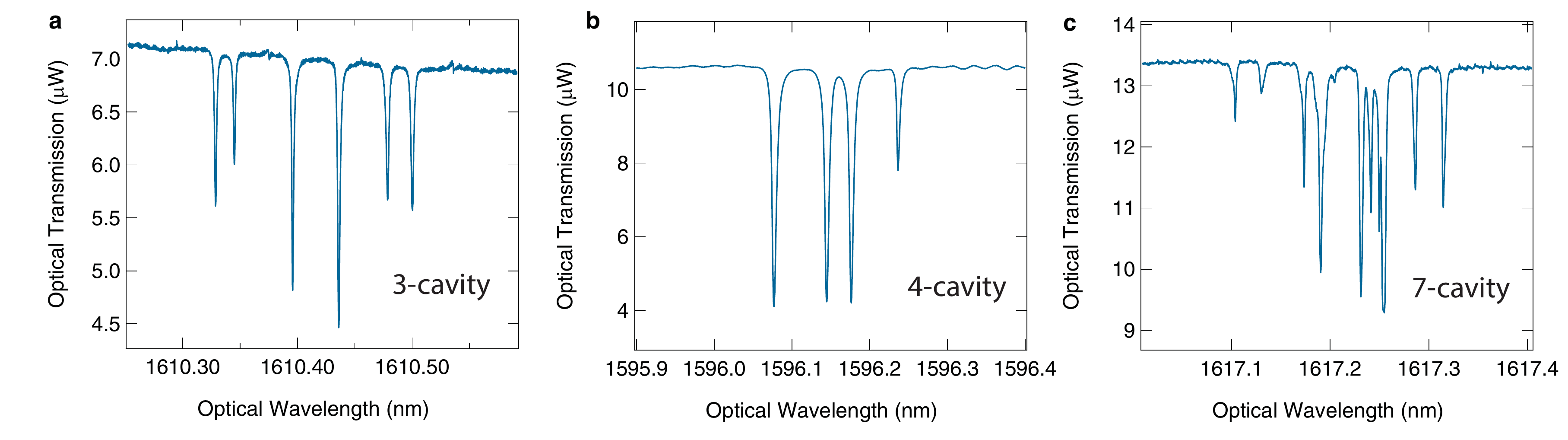}

\caption{The optical transmission in the (a) three,  (b)four and (c) seven cavity arrays respectively. The well split resonant mode and the high extinction of each split supermodes shows that the optical modes are strongly coupled.}
  \label{fig:opttrans}
\end{figure}
The coupled mode equation that governs the optical part reads
 
\begin{equation}
\begin{aligned}
\label{eq:opt_diagonal}
\left(\begin{array}{c}
\dot{a}_{1}\\
\dot{a}_{2}\\
...\\
\dot{a}_{i}\\
...\\
\dot{a}_{n}
\end{array}\right)&=
\left(
\begin{array}{cccccc}
 -\frac{\gamma _1}{2}-\mathrm{i} \omega _1 & \frac{\mathrm{i} \kappa_{12} }{2} & ... & \frac{\mathrm{i} \kappa_{1i} }{2} & ... & \frac{\mathrm{i} \kappa_{1n} }{2} \\
 \frac{\mathrm{i} \kappa_{21}}{2} & -\frac{\gamma _2}{2}-\mathrm{i} \omega _2 & ... & \frac{\mathrm{i} \kappa_{2i}}{2} & ... & \frac{\mathrm{i} \kappa_{2n} }{2} \\
 ... & ... & ... & ...& ... & ...\\
 \frac{\mathrm{i} \kappa_{i1} }{2} & \frac{\mathrm{i} \kappa_{i2} }{2} & ... & -\frac{\gamma _i}{2}-\mathrm{i} \omega _i& ... &  \frac{\mathrm{i} \kappa_{in} }{2}\\
 ... & ... & ... & ...& ... & ...\\
 \frac{\mathrm{i} \kappa_{n1} }{2} & \frac{\mathrm{i} \kappa_{n2} }{2} & ... & \frac{\mathrm{i} \kappa_{in} }{2}& ... & -\frac{\gamma _n}{2}-\mathrm{i} \omega _n\\
\end{array}\right)
\left(\begin{array}{c}
{a}_{1}\\
{a}_{2}\\
...\\
{a}_{i}\\
...\\
\dot{a}_{n}
\end{array}\right)
+\sqrt{\gamma _1 \eta _{c}}s(t) 
\left(\begin{array}{c}
1\\
0\\
...\\
0\\
...\\
0
\end{array}\right)
\end{aligned}
\end{equation}

where $\gamma_i, \omega_i$ are the total damping rate and the resonant frequency of each individual optical mode $a_i$. $\kappa_{ij}$ is the optical coupling strength between $i$th and $j$th OMO. $|s(t)|^2$ is the optical power that excites the array of resonators through the first cavity. $\eta_c$ is the optical criticality factor, the ratio between the cavity damping rate and the external coupling rate. For triangular lattice cases, the matrix dimension is increased from $n$ to $2n$ where the both the cw and ccw degrees of freedoms are taken into account. In the strong coupling regime ($\kappa_{ij}\gg\gamma_i ~\forall i,j$), individual optical modes hybridize and fully splits into optical supermodes $b_m$ where some supermodes span over all cavities. The spatial geometry of these supermodes can be analyzed through rewriting equation \ref{eq:opt_diagonal} in the optical supermode basis $b_m = \sum\limits_{i=1}^{N}c_i a_i$ where $c_i$ indicates the relative optical field amplitude in cavity $i$. To conveniently visualize the spatial geometry of the optical supermode, we approximate $\gamma_i$ and $\Omega_i$ of individual OMO being identical. We can then exactly diagonalize the coupling matrix in equation \ref{eq:opt_diagonal} through Jordan matrix decomposition to find the new optical supermodes basis $b_m$ in superposition of the individual optical mode $a_i$. The geometric representation of the superposition is shown in figure 2 in the main text. 

\begin{figure}
  \centering

  \includegraphics[angle=0,width=0.8\textwidth]{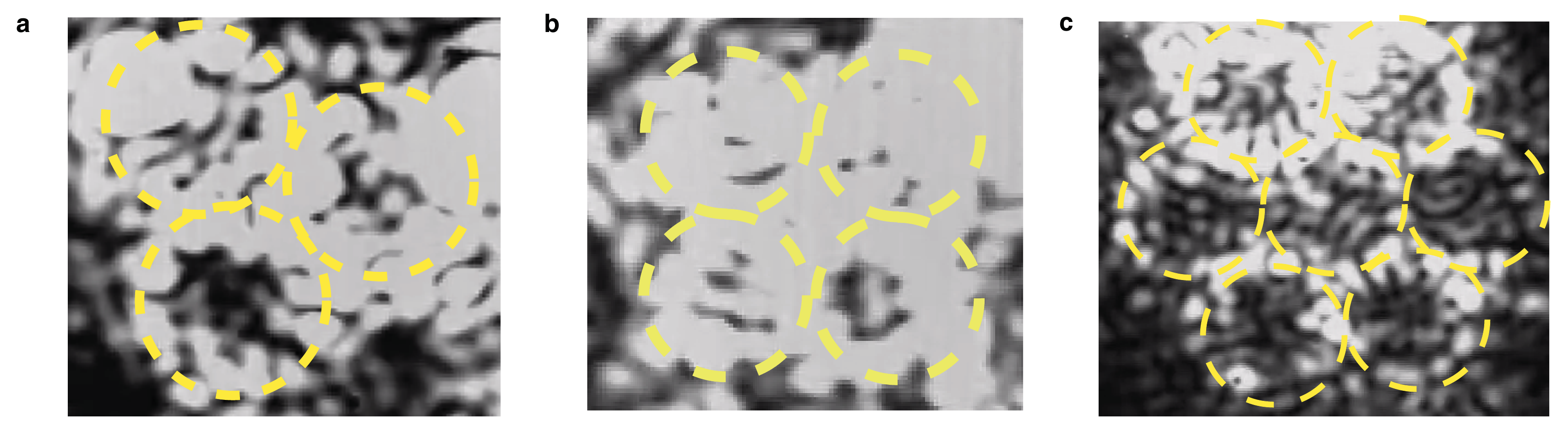}
\caption{Image of the scattered light when sychronization is onset for (a) three-cavity (b) four-cavity and (c) seven-cavity arrays. The scale bars are 20 $\mu$m.}
  \label{fig:mult_IR}
\end{figure}

Figure \ref{fig:mult_IR} shows images from the IR camera when the supermodes spanning over all cavities are excited. It is clear from the image that all cavities light up with approximately equal intensities. The strong coupling is also visualized by the fully split optical supermodes shown in the optical transmission in figure \ref{fig:opttrans}.

\subsection{Mechanical modes}
\begin{figure}
  \centering

  \includegraphics[angle=0,width=0.8\textwidth]{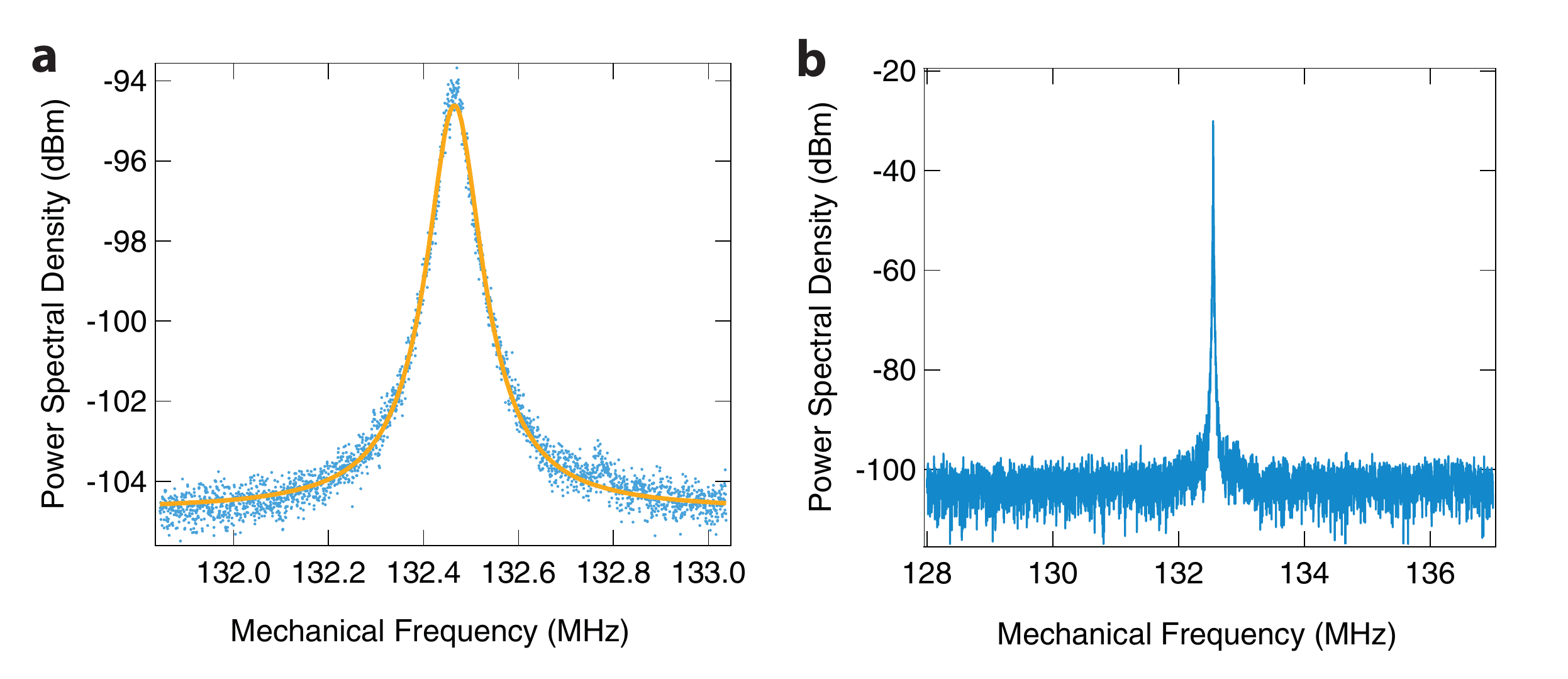}
\caption{(a) Thermal mechanical spectrum of a single doubled-disk OMO forming the array in this experiment. (b) Self-sustaining oscillation spectrum of a single double-disk OMO.}
  \label{fig:mult_mech}
\end{figure}

The mechanical modes of the double disk are the flapping mechanical mode where we can control the mechanical frequency of the mode by the undercut depth of the sacrificial silicon dioxide layer. In the particular device we used in the experiment, we fabricated the devices with mechanical frequencies near 132.5 MHz and quality factor near 1000 (Fig. \ref{fig:mult_mech}). Typical optical quality factor of these devices is $\sim$ 500,000. interaction strength between the optical and the mechanical mode is characterized by the optomechanical coupling strength $g_{\rm{om}}=\frac{d\omega}{dx}=2\pi\times 50 $ GHz/nm for our devices \cite{zhang_synchronization_2012}.

\begin{figure}[!h]
  \centering

  \includegraphics[angle=0,width=1\textwidth]{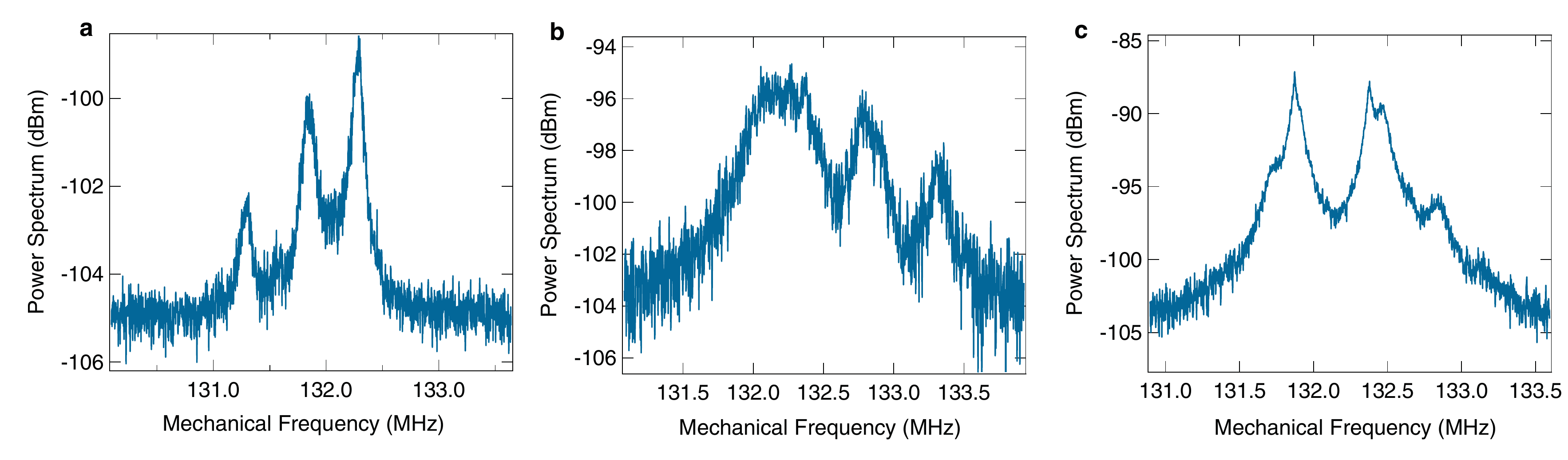}
\caption{The mechanical spectrum in the (a) three,  (b)four and (c) seven cavity arrays at low optical powers respectively. The disorder in mechanical frequencies is evident from the splitting of the peaks.}
  \label{fig:mult_mech2}
\end{figure}

In the OMO arrays, when the system is coupled to a laser at very low power, we measure a distribution of mechanical frequencies corresponding to different OMOs. Figure \ref{fig:mult_mech2} shows the mechanical spectrum in the three, four and seven cavity arrays respectively. In the three OMO array, the mechanical frequencies are clearly different whereas in the four and seven OMO arrays, the mechanical mode frequencies are different but also overlapping.

\subsection{Phase Noise}
We measure the phase noise of our oscillator by feeding the output of the photodiode (New Focus 1811) to a spectrum analyzer (Agilent E4407B). To measure the phase noise at 10 kHz offset frequency, we utilize the spot frequency measurement option in the phase noise measurement module (Option 226) of the spectrum analyzer which enables fast noise spectral power measurement at a single offset frequency. 

In synchronized oscillator arrays, the phase noise reduces as the number of synchronized oscillators increase. Experiments and theories \cite{chang_phase_1997,pikovsky_synchronization:_2003} show that the noise reduces as $1/N$ where $N$ is the number of synchronized oscillators. This can be intuitively understood as coordinated oscillators while having the same frequency but having an increased effective mass which improves their resistance to thermomechanical fluctuations. In our system, proving synchronization by studying the dynamics of individual cavities in the arrays is desirable but it is technically challenging given the compact size of the array. Therefore we choose the phase noise as a figure of merit as it is fundamentally limited by the thermomechanical process. 

\begin{figure}
  \centering
  \includegraphics[angle=0,width=0.5\textwidth]{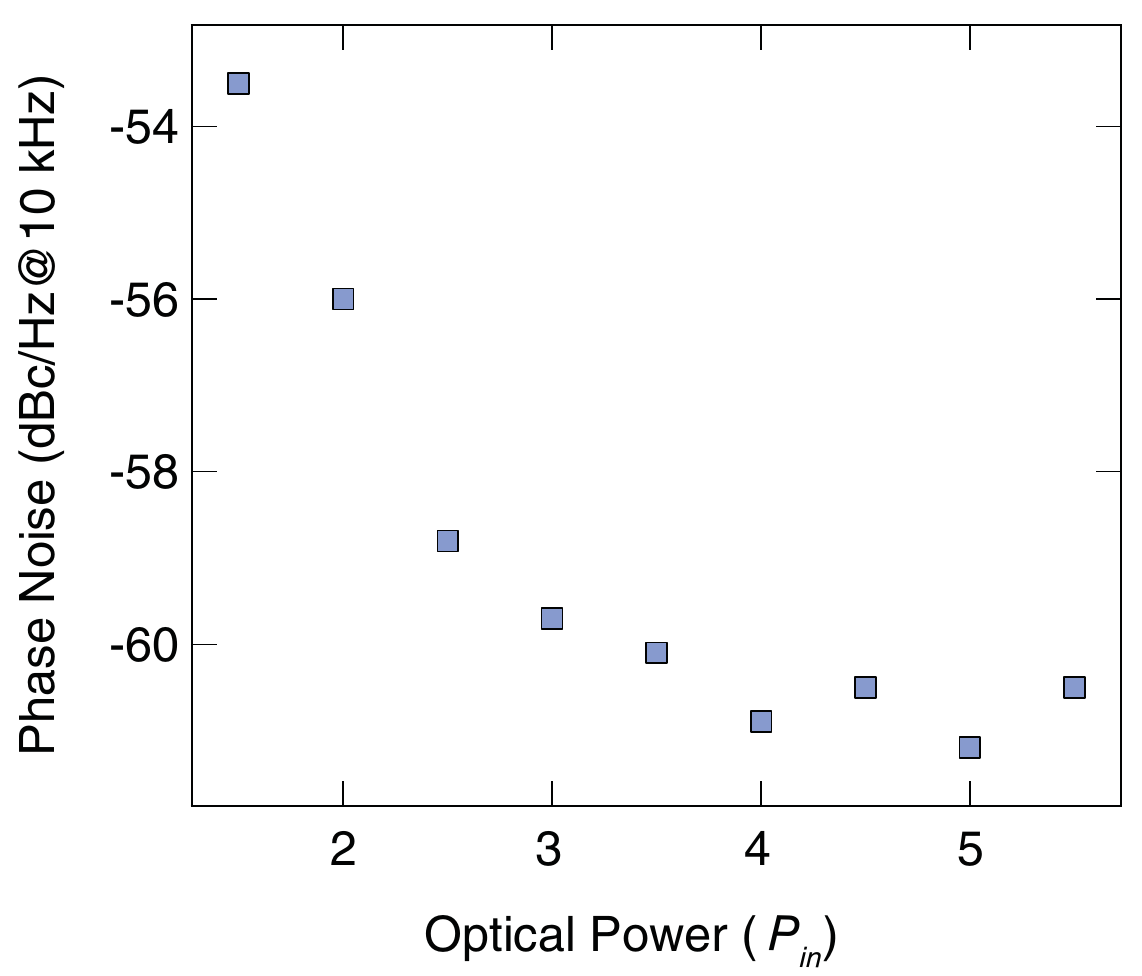}
\caption{Spot frequency phase noise of a single OMO at 10 kHz offset from the carrier frequency as a function of laser power. $P_{\rm{th}}=10 ~\mu$W is the threshold power for a single OMO to oscillate.}
  \label{fig:pn_one}
\end{figure}

The phase noise performance of an optomechanical oscillator, when limited by thermomechanical noise, improves when the laser power is increased. At high laser driving powers, the phase noise roughly stays constant as the laser power is further increased.  We calibrate the single oscillator phase noise performance by measuring the phase noise at 10 kHz carrier offset at different laser driving powers. Figure \ref{fig:pn_one} clear shows that the phase noise does not vary significantly when the power is above $4\times P_{th}$. where $P_{\rm{th}}=10 ~\mu$W is the typical threshold power for a single OMO. This is in agreement with the experimental observation reported by Mani et. al \cite {hossein-zadeh_characterization_2006}. and theory by Fong et. al \cite{fong_phase_2014}.


\begin{figure}
  \centering
  \includegraphics[angle=0,width=0.8\textwidth]{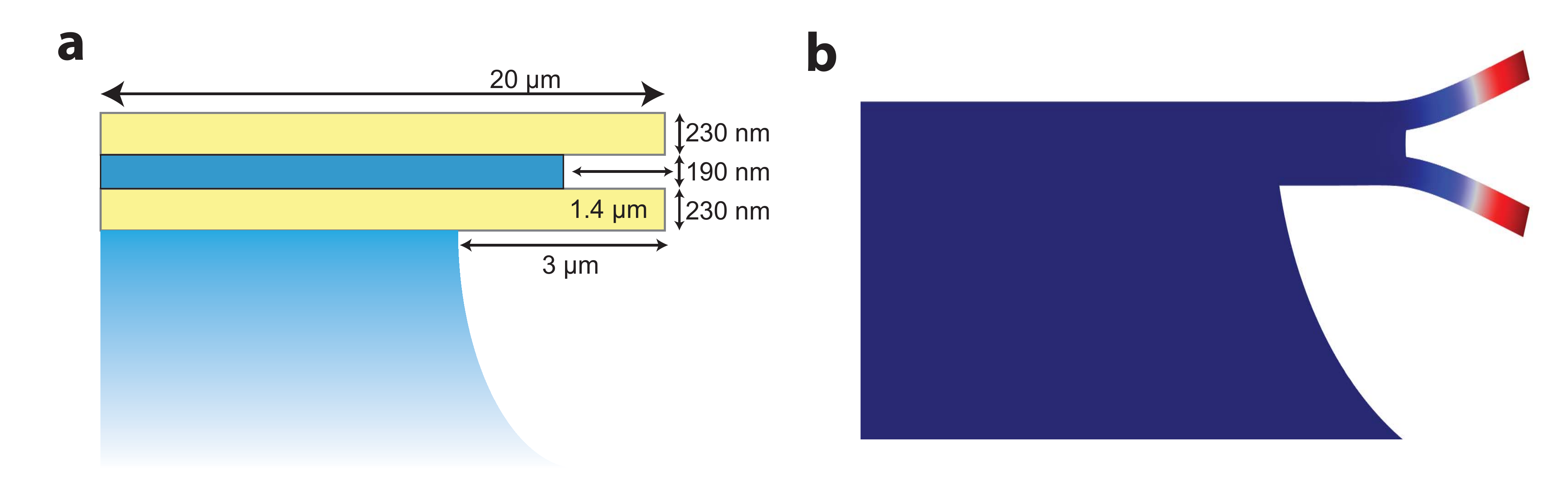}
\caption{FEM simulation of the mechanical mode. (a) Geometry of the structure. (b) Simulated mechanical displacement profile. The antisymmetric mechanical mode simulated has a vibration frequency of 131 MHz with an effective mass of 70 pg.}
  \label{fig:mech_sim}
\end{figure}

Leeson’s model \cite{leeson_simple_1966} predicts the phase noise of self-sustaining oscillators from the linewidth of the oscillation signal. The linewidth ($\Delta \Omega$) of a thermally limited oscillator is given by
\begin{equation}
\label{eq:linwid}
\Delta \Omega = \frac{k_B T}{2 P_\textnormal{out}}\Gamma_m^2
\end{equation}
where $k_B$ is Boltzmann's constant, $T$ is the temperature, $P_\textnormal{out}$ is the output power of the oscillator and $\Gamma_m$ is the natural damping rate of the oscillator. Since the double disk we use have a small effective mass $m_\textnormal{eff}$ (Fig. \ref{fig:mech_sim}), which means a low oscillator power, the phase noise is limited by thermomechanical noise. The oscillator power $P_\textnormal{out}$ is given by,

\begin{equation}
P_\textnormal{out} = \frac{1}{2}m_\textnormal{eff}\Omega^2 x^2 \Gamma_m
\end{equation}

where $x$ is the maximum displacement amplitude of the oscillator. Equation \eqref{eq:linwid} can be represented as,
\begin{equation}
\label{eq:linwidexp}
\Delta \Omega = \frac{k_B T}{ m_\textnormal{eff}\Omega^2 x^2}\Gamma_m
\end{equation}

In our experiment, since the offset frequency $\Delta f$ we are interested in is much larger than the linewidth ($\Delta \Omega$) of our oscillator, the phase noise $L(\Delta f)$ in dBc/Hz predicted by Leeson's equation\cite{leeson_simple_1966} can be simplified as,

\begin{equation}
\label{eq:lee}
L(\Delta f)=10 \textnormal{Log}_{10}\left(\frac{\Delta \Omega}{2\pi \Delta f^2}\right)
\end{equation}

Substituting the parameters ($T=300$ K; $m_\textnormal{eff}=70$ pg; $\Gamma_m=2\pi \times 110$ kHz) and $x\sim g_{om}/\gamma = 10$ pm to equation \ref{eq:lee}, we obtain a theoretical estimate of the thermomechanical noise limit of our oscillators which is $-60$ dBc/Hz at $10$ kHz carrier frequency offset in good agreement with our experimental measurements and previous frequency dependent phase noise measurements \cite{shah_master-slave_2015}. 

We expect the laser noise to be a minor contribution to the phase noise of our OMOs at 10 kHz carrier offset. Although diode lasers typically have relatively high phase noise, at this frequency, the phase noise of the laser is not relevant because the optical force is only affected by laser intensity. The cavity decay rate ($\frac{\gamma}{2\pi}\sim 400$ MHz) is much higher than 10 kHz which means the laser phase noise does not get converted in to intensity noise due to the cavity response. Typical relative intensity noise (RIN) of external cavity tunable laser at low frequencies is below -100 dBc/Hz \cite{laurila_tunable_2002}. We measure the RIN of the laser we use (Tunics Reference) at 50 $\mu$W to be -112 dBc/Hz at 10 kHz offset through the photodiode (New Focus 1811) and the spectrum analyzer (Agilent E4407B). As laser shot noise is much lower than technical noise at this laser power and frequency, we expect minimal contribution from the shot noise of the laser. Our analysis is consistent with previous theory \cite{fong_phase_2014}. \\

\end{document}